\documentclass[acmsmall]{acmart}

\usepackage{color}
\usepackage{url}
\usepackage{multirow}
\usepackage{enumerate}
\usepackage{graphicx}
\usepackage{braket}
\usepackage{todonotes}
\usepackage{amsmath}
\usepackage{hyperref}
\usepackage{amsmath}   
\usepackage{mathrsfs}
\usepackage{amsfonts}
\usepackage{mathtools}
\usepackage{comment}
\usepackage{longtable}
\usepackage{multicol}

\usepackage{wrapfig}
\usepackage{lscape}
\usepackage{rotating}
\usepackage{booktabs}
\usepackage[ruled,vlined]{algorithm2e}
\DeclareUnicodeCharacter{2212}{-}

\usepackage{mathtools}

\SetCommentSty{mycommfont}
\SetKwInput{KwInput}{Input}                
\SetKwInput{KwOutput}{Output} 
\SetKw{Continue}{continue}
\SetKw{Break}{break}
\newcommand{\bX}{\mathbf{X}}
\newcommand{\bx}{\mathbf{x}}

\begin{document}

\title{$i$-QER: An Intelligent Approach towards Quantum Error Reduction}

\author{Saikat Basu}
\email{saikat.basu.000@gmail.com}
\author{Amit Saha}
\email{abamitsaha@gmail.com}
\affiliation{
  \institution{A. K. Choudhury School of Information Technology, University of Calcutta}
  \city{Kolkata}
  \country{India}
  \postcode{700106}
}
\affiliation{
  \institution{Atos, Pune}
  \country{India}
  \postcode{411045}
}

\author{Amlan Chakrabarti}
\affiliation{%
  \institution{A. K. Choudhury School of Information Technology, University of Calcutta}
  \city{Kolkata}
  \country{India}
  \postcode{700106}
}
 
 \author{Susmita Sur-Kolay}
\affiliation{%
  \institution{Advanced Computing \& Microelectronics Unit, Indian Statistical Institute}
  \city{Kolkata}
  \country{India}
  }

\begin{abstract}
  Quantum computing has become a promising computing approach because of its capability to solve certain problems, exponentially faster than classical computers. A $n$-qubit quantum system is capable of providing $2^{n}$ computational space to a quantum algorithm. However, quantum computers are prone to errors. Quantum circuits that can reliably run on today's Noisy Intermediate-Scale Quantum (NISQ) devices are not only limited by their qubit counts but also by their noisy gate operations. In this paper, we have introduced $i$-QER, a scalable machine learning-based approach to evaluate errors in a quantum circuit and helps to reduce these without using any additional quantum resources. The  $i$-QER predicts possible errors in a given quantum circuit using supervised learning models. If the predicted error is above a pre-specified threshold, it cuts the large quantum circuit into two smaller sub-circuits using an error-influenced fragmentation strategy for the first time to the best of our knowledge. The proposed fragmentation process is iterated until the predicted error reaches below the threshold for each sub-circuit. The sub-circuits are then executed on a quantum device. Classical reconstruction of the outputs obtained from the sub-circuits can generate the output of the complete circuit. Thus, $i$-QER also provides classical control over a scalable hybrid computing approach, which is a combination of quantum and classical computers. The $i$-QER tool is available at \href{https://github.com/SaikatBasu90/i-QER}{https://github.com/SaikatBasu90/i-QER}.    
  
\end{abstract}

\maketitle

\section{Introduction}

With the progress of quantum computing in the last two decades, modern researchers have exhibited a wondrous enthusiasm for realising quantum algorithms \cite{chuang, grover, Shor_1997}  to attain asymptotic improvement. Physical quantum computers employing various technologies, such as  continuous spin systems \cite{Preskill2018quantumcomputingin}, superconducting transmon technology \cite{PhysRevA.76.042319}, nuclear magnetic resonance \cite{Dogra_2014, Gedik_2015}, photonic systems \cite{Slussarenko_2019},
 ion trap \cite{Bruzewicz_2019}, topological quantum systems \cite{Cui_2015first}, have reached new heights recently. It has become a blazing domain among the researchers to implement quantum algorithms  \cite{Preskill2018quantumcomputingin} on the physical quantum devices. Every quantum algorithm can be realized in the form of a quantum circuit through qubits and quantum gates. Albeit the colossal promises of quantum algorithms, the present quantum computers are more error-prone, which limits the capability of solving a computation problem in quantum devices.

 Quantum error correcting codes (QECCs) \cite{10.1007/3-540-46796-3_23, 10.1007/3-540-49208-9_27, Baek_2019,Leuenberger_2001, PhysRevA.52.R2493, PhysRevLett.77.793, Wootton_2020, RevModPhys.87.307, PhysRevLett.77.198} are used to eradicate errors that arise from noise, to provide an avenue toward fault-tolerant quantum computation. However, in  practice, the implementation of quantum error correction imposes a huge burden in terms of the required number of qubits, which remains beyond the capabilities of present near-term devices. Due to the typical error rate of current near-term devices, various error reduction techniques have emerged; one of them is quantum error mitigation (QEM) \cite{Nautrup_2019,9226505}. QEM does not use any extra quantum resources, rather it aims to slightly enhance the accuracy of estimating the outcome in a given quantum computational problem through several techniques such as extrapolation, probabilistic error cancellation, quantum
subspace expansion, symmetry verification, machine learning etc. \cite{Nautrup_2019}. As per the state-of-the-art work, QEM is restricted to quantum circuits having a very limited number of qubits and limited depth due to the enormous overhead of classical computational time complexity. 

In order to overcome the engineering challenges of QEM, fragmentation of quantum circuits can be a good approach because it breaks up quantum circuits into smaller sub-circuits or partitions, with fewer qubits and shallower depth. Thus the sub-circuits have to deal with short coherence times of noisy intermediate-scale quantum processor. Each sub-circuit faces a lower effect of noise when executed in a NISQ device. Primarily, fragmentation of a quantum circuit is used to simulate a larger quantum system on a small quantum computer \cite{Bravyi_2016, PhysRevLett.125.150504}. In \cite{10.1145/3445814.3446758, 9155024}, the authors have proposed fragmenting a quantum circuit in such a way that the exponential post processing cost can be optimized. Further, in \cite{ayral2021quantum}, the authors explored  for the first time how such fragmentation affects the various quantum noise models. Later in \cite{Perlin2021QuantumCC}, it was explored that circuit cutting helps to mitigate the effects of noise as portrayed in \cite{ayral2021quantum}. Although, the principle objective of the previous works were to make the large circuit implementable by fragmenting them, but the noise leading to inaccurate outcomes was never addressed.

\textbf{Our motivation:} With this background, it is evident that fragmentation of a quantum circuit into smaller sub-circuits can  reduce the effect of noise when executed in a NISQ device. Nonetheless, fragmentation increases post-processing cost, and thereby overall computational cost also escalates. In this paper, error reduction of a given circuit is carried out by predicting error in a quantum circuit, keeping into account that the fragmentation of a given quantum circuit is minimized. However, predicting error in a quantum circuit is a non-trivial problem. In this article, a machine learning based approach has been adopted for better accuracy in predicting the error in a quantum circuit by training the system considering the features of a quantum circuit. Our main contributions are as follows:
\begin{itemize}
\item a supervised machine learning based prediction technique to evaluate errors in a given quantum circuit;

\item an error influenced fragmentation strategy called \textit{Error Influenced Binary Quantum Circuit Fragmentation} based on the proposed prediction system has been introduced, to reduce the error of a quantum circuit; and

\item an automated machine learning based tool $i$-QER i.e., \textit{An Intelligent Approach towards Quantum Error Reduction} that helps to reduce the error of a quantum circuit that can be further incorporated for any existing quantum hardware. This tool also provides a classical control over hybrid quantum-classical computing.
\end{itemize}

The structure of this paper is as follows. Section 2 briefly presents the preliminary concepts of quantum circuits, quantum hardware, quantum circuit fragmentation and supervised machine learning models. Section 3 proposes the automated machine learning based approach $i$-QER. Section 4 briefly discusses about the experimental results of the proposed methodology. Section 5 captures our conclusions.



\section{Background}

\subsection{Quantum Circuit}

A quantum circuit is the schematic representation of any quantum algorithm or quantum program. Each line in the quantum circuit is expressed as a qubit and the operations, i.e., quantum gates are illustrated by different blocks on the line \cite{PhysRevA.52.3457}. In Table \ref{tab:gate}, we summarize the commonly used logical quantum gates.

\begin{table}[!ht]
\centering
\caption{Matrix Representation of Quantum Gates}
\scriptsize

\begin{tabular}{  c | c | c  }
 \hline
 Operator & Quantum Gates & Matrix Representation \\
 \midrule
 Hadamard & \includegraphics[scale=.5]{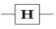} & $\begin{pmatrix}
\\ \frac{1}{\sqrt{2}} & \frac{1}{\sqrt{2}}
\\ \frac{1}{\sqrt{2}} & -\frac{1}{\sqrt{2}}
\end{pmatrix}$ \\
 \hline


Pauli-X & \includegraphics[scale=.5]{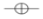} &  $\begin{pmatrix}
 \\ 0 & 1
 \\ 1 & 0
 \end{pmatrix}$ \\
 \hline
 Pauli-Y & \includegraphics[scale=.5]{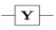} & $\begin{pmatrix}
 \\ 0 & -i
 \\ i & 0
 \end{pmatrix}$ \\
 \hline
  Pauli-Z & \includegraphics[scale=.5]{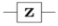} & $\begin{pmatrix}
 \\ 1 & 0
 \\ 0 & -1
 \end{pmatrix}$ \\
 \hline
 T & \includegraphics[scale=.5]{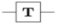} & $\begin{pmatrix}
 \\ 1 & 0
 \\ 0 & e^{i\pi/4}
 \end{pmatrix}$ \\
 \hline
 S & \includegraphics[scale=.5]{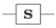} & $\begin{pmatrix}
 \\ 1 & 0
 \\ 0 & i
 \end{pmatrix}$ \\
 \hline

CNOT & \includegraphics[scale=.5]{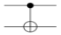} & $\begin{pmatrix}
\\ 1 & 0 & 0 & 0
\\ 0 & 1 & 0 & 0
\\ 0 & 0 & 0 & 1
\\ 0 & 0 & 1 & 0

\end{pmatrix}$ \\
\hline
 TOFFFOLI & \includegraphics[scale=.5]{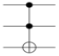} & $\begin{pmatrix}
\\ 1 & 0 & 0 & 0 & 0 & 0 & 0 & 0
\\ 0 & 1 & 0 & 0 & 0 & 0 & 0 & 0
\\ 0 & 0 & 1 & 0 & 0 & 0 & 0 & 0
\\ 0 & 0 & 0 & 1 & 0 & 0 & 0 & 0
\\ 0 & 0 & 0 & 0 & 1 & 0 & 0 & 0
\\ 0 & 0 & 0 & 0 & 0 & 1 & 0 & 0
\\ 0 & 0 & 0 & 0 & 0 & 0 & 0 & 1
\\ 0 & 0 & 0 & 0 & 0 & 0 & 1 & 0
\end{pmatrix}$ \\

 \midrule
\end{tabular}
\label{tab:gate}
\end{table}

\subsection{Quantum Hardware}
\begin{figure}[!ht]
\centering
\includegraphics[width=50mm, height=2cm]{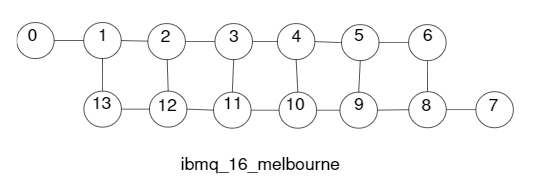}
\caption{Qubit Topology \cite{IBM}}
\label{qt}
\end{figure}

 Currently, the most popular quantum technologies for implementing the quantum gates are superconducting quantum devices, quantum dots, ion traps, neutral atoms. For mapping the  synthesized quantum logic circuit to an existing noisy quantum hardware or NISQ (Noisy Inter-mediate Scale quantum) device \cite{Preskill2018quantumcomputingin}, the logical quantum gates described in Table \ref{tab:gate}, must be realized by technology-specific gates to be hardware compatible for implementation. Each one of the devices based on the above mentioned technologies has a specifically dedicated qubit topology. For example, the arrangement for the 15-qubit IBM Melbourne quantum hardware \cite{IBM} is  shown in Fig. \ref{qt}. All the experiments of this paper are carried out on this machine. 

Along with the constraint on the number of qubits, NISQ devices are “noisy”. Thus, estimation of the simulated result of a quantum state is erroneous. This error has to be reduced as much as possible, to get a proper estimation of the simulated result of a quantum state closer to its expected outcome.

\subsection{Quantum Circuit Fragmentation and Output Reconstruction}
\label{section fragment}

 Fragmentation of a quantum circuit can be defined as a method for simulating large circuits by partitioning them into multiple partitions of smaller circuits which are mutually interacting weakly. In this article, our main aim is to reduce the circuit error through fragmentation of a quantum circuit.

\begin{figure}[!ht]
\centering
\includegraphics[width=140mm, height=5.5cm]{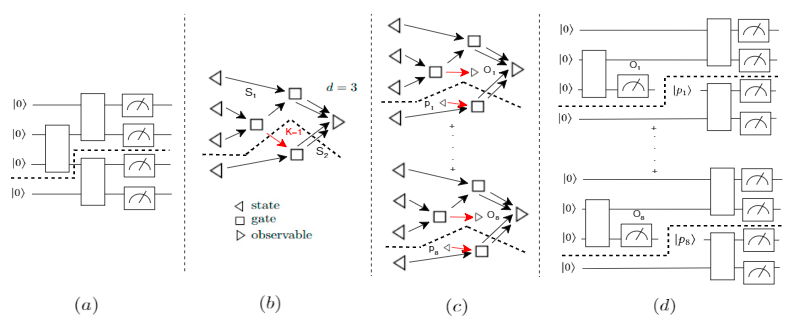}
\caption{Four phases of quantum circuit fragmentation method as described in \cite{PhysRevLett.125.150504}: (a) an example quantum circuit; (b) its tensor network; (c) a
collection of tensor networks obtained by cutting an edge; and (d) fragmented smaller quantum circuits for the example circuit. }
\label{frag}
\end{figure}

   Four well-defined phases of the circuit fragmentation method are shown in Fig. \ref{frag} \cite{PhysRevLett.125.150504}. Let us consider an example 4-qubit circuit $C$ with three two-qubit gates as shown in Fig. \ref{frag}(a). All the input qubits are initialized with $\ket{0}$ and all the output qubits are measured in the same computational basis. Now, $C$ can be described by a tensor network $(G, A)$ consisting of a directed graph $G(E,V)$ and a collection of tensors $A=\{A(v): v \in V\}$. In Fig. \ref{frag}(b), the corresponding tensor network of $C$ is shown. As per conventional representation of a tensor network, the vertices $V$ can be represented by individual gates (denoted by~$\Box$), input qubits (denoted by~$\lhd$), and measurement operators (denoted by~$\rhd$) as shown in  Fig. \ref{frag}(b), considering that the flow of qubits is encoded by the directed edges $E$. For each $v \in V$, $A(v)$ is a tensor that encodes the matrix entries of the corresponding gate, state, or measurement operator, and the value $T(G, A)$ of the   tensor network $(G, A)$ occurs as the expected output of the corresponding circuit $C$. As shown in  Fig. \ref{frag}(b), the two parts of a partition $\{S_1, S_2\}$ are indicated by a dashed line. Considering  that only one qubit, i.e., the third qubit is sent from $S_1$ to $S_2$ as indicated by the red arrow,  the value of $K$ is 1, where $K$ can be defined as the number of edges between the two partitions of $G$. Due to two outgoing edges from $S_1$ to measurement operators and one from $S_1$ to $S_2$, the value of $d(S_1)$ is 3, where $d$ can be defined as the number of qubits sufficient for simulating each partition. Similarly, the value of $d$ for the partition of $S_2$ is 2 and therefore we can conclude that $d$ is 3, by considering $max(d(S_1), d(S_2))$. On that account, the example circuit $C$ can be designated as $(1,3)$-partitioned, as the values of $K$ and $d$ are 1 and 3 respectively. For simulating the two partitions on a $3$-qubit quantum hardware, a well-defined edge-cutting procedure to realize the tensor networks as described in \ref{frag}(b), is shown in Fig. \ref{frag}(c). After cutting an edge $e \in E$, the tensor $T(G(E,V),A)$ for the tensor network $(G(E,V),A)$ of the example circuit $C$ can be mathematically represented as,
\begin{equation} \label{eqn:frag}
  T(G, A) = \sum_{i=1}^8 c_i T(G',A_i),
\end{equation}
where $G'$ differs from $G$ by removing $e$ and adding one $\lhd$ and one $\rhd$ vertex, each $c_i \in \set{-\frac{1}{2},\frac{1}{2}}$, and each $(G', A_i)$ corresponds to a valid quantum circuit. This can be further expressed in terms of intermediate measurement operator $O_i$ and states $p_i$, where $1\le i \le 8$. Thus, the tensor network is partitioned into two sub-circuits, which does not eventually affect the value of the overall tensor network. Each partition must have outgoing edges to the measurement operator and can hence be simulated by a $3$-qubit quantum computer followed by classical processing of the measurement outcomes, as shown in Fig. \ref{frag}(d).

Here, the individual simulation results are combined by a simple sampling procedure according to Equation \ref{eqn:frag} where each Pauli matrix $\{I, X, Y, Z\}$ is expanded in its eigenbasis as denoted by $O_i$ and their eigenprojectors  $p_i$. Thus mathematically, $\mathbf{A}$, an arbitrary 2$\times$2 matrix, can be realized as
\begin{equation}
    \mathbf{A}=\frac{Tr(\mathbf{A}I)I+Tr(\mathbf{A}X)X+Tr(\mathbf{A}Y)Y+Tr(\mathbf{A}Z)Z}{2}.
\end{equation}

For execution on quantum computers, the Pauli matrices are further expressed in terms of their eigenbases as follows
\begin{equation}
    \mathbf{A}=\frac{A_1+A_2+A_3+A_4}{2}
\end{equation}
where
\begin{eqnarray*}
    A_1&=&[Tr(\mathbf{A}I)+Tr(\mathbf{A}Z)]\ket{0}\bra{0}\\
    A_2&=&[Tr(\mathbf{A}I)-Tr(\mathbf{A}Z)]\ket{1}\bra{1}\\
    A_3&=&Tr(\mathbf{A}X)[2\ket{+}\bra{+}-\ket{0}\bra{0}-\ket{1}\bra{1}]\\
    A_4&=&Tr(\mathbf{A}Y)[2\ket{+i}\bra{+i}-\ket{0}\bra{0}-\ket{1}\bra{1}]
\end{eqnarray*}

Each trace operator corresponds physically to measure the qubit in one of the Pauli bases $O_i \in \{I, X, Y, Z\}$ and each of the density matrices corresponds physically to initialize the qubit in one of the eigenstates $p_i\in\{\ket{0},\ket{1},\ket{+},\ket{+i}\}$. 

Recently, in \cite{Perlin2021QuantumCC}, maximum-likelihood fragment tomography (MLFT) was introduced as an improved circuit cutting technique, with a limited number of qubits, to run partitioned quantum sub-circuits on quantum hardware. MLFT further finds the most likely probability distribution for the output of a quantum circuit, with the measurement data obtained from the circuit’s fragments, along with minimizing the classical computing overhead of circuit cutting methods. Hence, they showed that circuit cutting as a standard tool can be used for running partitioned sub-circuits on quantum devices by estimating the outcome of a partitioned circuit with higher fidelity as compared to the full circuit execution. In this work, we have used a supervised machine learning based error influenced fragmentation strategy to cut the circuit into two partitions,  described in next section.

\subsection{Supervised Machine Learning}

Machine Learning is the development of computing systems capable of learning automatically, without following explicit instructions. It uses algorithms or statistical models to extract inferences from patterns in the data \cite{10.5555/1734076}. Supervised learning algorithm is one of the key approaches of machine learning, which builds a mathematical model from a labelled data-set. The labelled data is known as training data, and consists of a set of training examples. Each training example consists of multiple feature values and the corresponding output label. The supervised learning algorithms used in this paper are described next.

\subsubsection{Linear Regression:} This is a linear approach to model the relationship between a scalar response and one or more explanatory variables. In most linear regression models, given a training set of ($x_i, y_i$), $i = 1$, $\dots$ , $N$, where $x_i \in R^n$ and $y \in \{1, −1\}^N$, the objective is to minimize the sum of squared errors,
\vspace{-0.4cm}
\begin{align}
    minimize & \;\;  \sum\limits_{i=1}^N (y_i - w_i x_i)^2
    \end{align}

where $w_i$ represents the coefficients, or the weights of linear regression model. Linear regression is a special case of polynomial regression, which is also used in this work. Polynomial regression converts the original independent variables ($x$) of training data into polynomial features of required degree ($2$, $3$, \dots, $n$). Therefore, it helps to make use of linear regression to train the more complicated non-linear data-set and increase the accuracy of the model.  

\subsubsection{Lasso Regression (Least Absolute Shrinkage and Selection Operator):} This is a special type of linear regression that uses shrinkage \cite{Tibshirani1996RegressionSA}  where data values are shrunk towards a central point, say the mean. The lasso procedure encourages simple sparse models. Given a data-set of ($X^i, y_i$), $i = 1$, $\dots$ , $N$, where $X^i= (x_{i1}, \dots, x_{ip})^T$ are the predictor variables and $y \in \{1, −1\}^N$ are the responses, then the objective is to estimate Lasso ($\hat{\alpha}, \hat{\beta}$),
\vspace{-0.25cm}
\begin{align}
    minimize & \;\;  \sum\limits_{i=1}^N (y_i - \alpha - \sum\limits_{j} \beta_j x_j)^2\\
    subject \: to & \;\; |\beta_j| \le t
\end{align}

where $t \ge 0$ is a tuning parameter,  $\hat{\beta}=\{\hat{\beta_1}, \dots, \hat{\beta_p}\}^T$, and $\alpha$ is the parameter which balances the amount of emphasis given to minimization object. While $\alpha=0$, Lasso regression produces the same coefficients as a linear regression.

\subsubsection{Support Vector Regression:}
The goal of a support vector machine (SVM) \cite{vapnik95} is to produce a model based on the training data, which predicts the target values of the test data given only the test data attributes. SVM can also be used as a regression method, i.e., support vector regression (SVR), employing the same principles as the SVM for classification, with only a few minor differences. First of all, as the output is a real number it becomes very difficult to predict the information in hand, which has infinite possibilities. A margin of tolerance ($\xi$) is set in approximation to the SVM. Although the main goal is always the same, i.e., to minimize error, here the hyperplane which maximizes the margin $(W, b)$, keeping in mind that part of the error is tolerated, is to be found. Given a training set of ($x_i, y_i$), $i = 1$, $\dots$ , $N$ where $x_i \in \mathbb{R}^n$ and $y \in \{1, −1\}^N$, the SVR requires the solution to the following optimization problem:

\begin{align}
    minimize & \;\; {1\over2} W^T W + C \sum\limits_{i=1}^N (\xi_i + \xi_i^*)\\
    subject \: to & \;\; y_i - W^T x_i - b \le \xi_i + \xi_i^*,\\
& -y_i + W^T x_i + b \le \xi_i + \xi_i^*, \\
& \;\; \xi_i , \xi_i^* \geq  0
\end{align}

where $C>0$ is the penalty parameter of the error term.

The SVM is trained using a non-linear support vector regressor. A kernel function has to be chosen to implement support vector regression. The kernel function takes the original non-linear problem and transforms it into a linear one in the higher-dimensional space. One of the popular kernel functions is Radial basis function (RBF). The RBF kernel estimates the similarity between two points by non-linearly mapping them to the higher dimensional space\cite{article}. Thus, it is capable of performing better than linear kernels in the presence of nonlinear attributes. Let us assume two points $x_{1}$ and $x_{2}$, represented as a feature vector in a training data. Then, the kernel can be mathematically represented as 

\begin{equation}
    K(x_{1},x_{2})=exp  \{- \frac{ \|x_{1}-x_{2}\|^{2}}{2 \sigma^{2}}\}= exp  (- \gamma \|x_{1}-x_{2}\|^{2})
\end{equation}
assuming $\gamma=1/2\sigma^{2}$. Note that $\|x_{1}-x_{2}\|^{2}$ is the euclidean distance between $x_{1}$ and $x_{2}$, and $\sigma$ is the variance.

\subsubsection{Random Forest Regression:}  
A random forest is a predictor consisting of a collection of randomized base regression trees $\{r_n(\bx, \Theta_m,\mathcal D_n), m\geq 1\}$, where $\Theta_1, \Theta_2,\hdots$ are independent and identically distributed outputs of a randomizing variable $\Theta$. These random trees are combined to form the aggregated regression estimate
\begin{equation}
\bar r_n(\bX, \mathcal D_n)=\mathbb E_{\Theta}\left[r_n(\bX,\Theta, \mathcal D_n)\right]
\end{equation}
where $\mathbb E_{\Theta}$ denotes the expectation with respect to the random parameter, conditionally on $\bX$ and  $\mathcal D_n$, a training data-set $\mathcal D_n=\{(\bX_1,Y_1), \hdots, (\bX_n,Y_n)\}$ of independent and identically distributed random variables.

\section{Proposed Framework for \lowercase{$i$}-QER}
\begin{figure}[]
  \centering
  \includegraphics[scale=0.4]{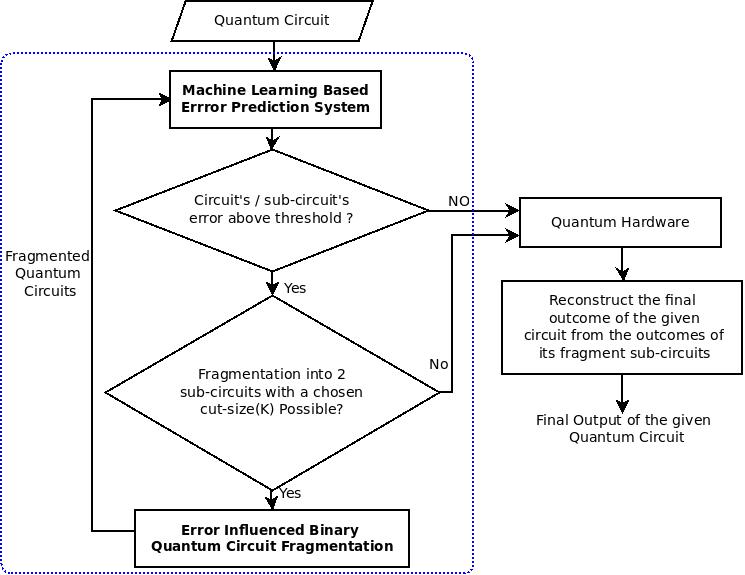}
  \caption{ An overview of $i$-QER, an intelligent approach towards quantum error reduction using quantum circuit fragmentation. The blue dotted box contains the basic building blocks of the tool.
  }
  \label{Block diagram}
\end{figure}

The details of our quantum error reduction tool $i$-QER are presented below. The tool takes a quantum circuit as an input, and a supervised learning based error prediction system predicts the possible effects of noise on the quantum circuit when it is executed on a Noisy Intermediate-Scale Quantum (NISQ) device  \cite{Preskill2018quantumcomputingin}. If the predicted effect of noise is beyond a threshold specified by the user, the circuit is fragmented \cite{PhysRevLett.125.150504} into two sub-circuits using a novel \textit{Error Influenced Binary Quantum Circuit Fragmentation} strategy. The same process is repeated till the circuit is ready to be executed within an acceptable error rate. The sub-circuits are then executed on quantum hardware. The outcome of the full circuit is constructed from the outputs (probability distributions) of the sub-circuits. The  output reconstruction of the full circuit can be adopted from \cite{PhysRevLett.125.150504, Perlin2021QuantumCC}, which is already described in Section 2.3. Fig. \ref{Block diagram} provides a flowchart of the tool incorporating the key components, of which our main contributions are indicated in bold.  

\subsection{Machine Learning Based Error Prediction System}


\subsubsection{Training Data-set Construction} 
For any prediction model, a good training data on which the model can learn is mandated. The prediction system learns from the training data collected from the outputs of a known set of circuits run on a particular quantum hardware. Any NISQ device executes a quantum circuit multiple times (shots). After each shot, all the qubits are measured and a classical output is recorded. After all the shots, a probability distribution over the observable states is returned. For a different number of shots, the output probability distribution are likely to be different. In order to construct the training data with consistent statistical error, we have generated it with a fixed number of shots for all the circuits considered. 

The data-set must contain the information about (i) the errors that occurred due to the noisy quantum computation, (ii) all the major features of a quantum circuit on which the error may be dependent and their corresponding errors that occurred while executing them on a NISQ device. 

\begin{figure}[h!]
  \centering
  \includegraphics[scale=0.34]{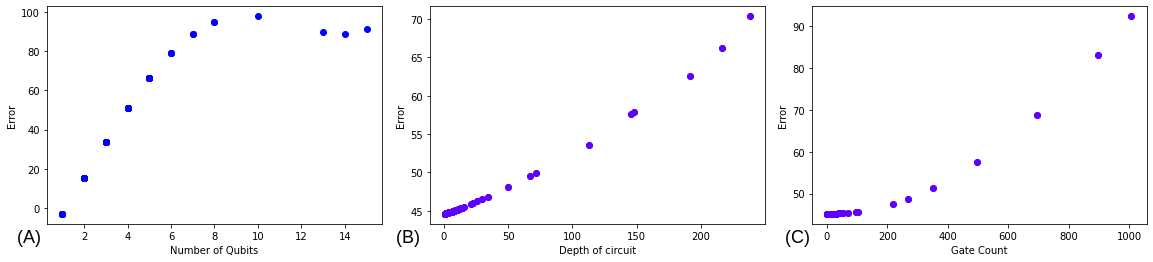}
  \caption{Effect on error(\%) in a circuit of its: (A)number of qubits, (B) depth, and (C) gate count.}
  \label{feature}
\end{figure}

The dependencies of error on various circuit parameters are shown in Fig. \ref{feature}. It shows that the impact of noise on a quantum circuit depends on a few major features, namely the number of qubits, the depth and the number of different quantum gate operations of the circuit.
 Hence, we need to extract the feature set for the training data based on these key components.
 The ideal measurement probabilities at a specific shot are compared with the actual measurement probabilities obtained under realistic noisy quantum hardware to obtain the dependent variable of the training data. We can describe probability vectors $P^{s}_{I}$ and $P^{s}_{N}$ to denote the ideal measurement probability and the actual noisy quantum hardware measurement probability respectively at a specific shot $s$. The error vector can be defined as $E^{s}=P^{s}_{N} - P^{s}_{I}$. 
 \begin{table*}[!ht]
\footnotesize
\centering
\caption{Independent features of the constructed data-set for a given circuit}
\resizebox{14cm}{!}{%
\begin{tabular}{c|c | c | c}

Circuit & \# Qubits & Depth & \# Gates of each of these 17 types \\
    \midrule
   & & & $H $ $\lvert$  $CNOT$ $\lvert$ $X$ $\lvert$ $Y$ $\lvert$ $Z$ $\lvert$ $R_{X}$ $\lvert$ $R_{Y}$ $\lvert$ $R_{Z}$ $\lvert$ $CZ$ $\lvert$ $CP/CU1$ $\lvert$  $T$ $\lvert$ $Toffoli$ $\lvert$ $SWAP$ $\lvert$ $T\dagger$ $\lvert$ $S$ $\lvert$ $S\dagger$ $\lvert$ $U3$  
\end{tabular}}
\label{tab:dataset}
\vspace{0.25cm}
\end{table*}

The features of the data-set used is specified in Table~\ref{tab:dataset}.  
The complete data-set constructed thus can be found at \href{https://github.com/SaikatBasu90/i-QER/blob/main/Prediction Model/QData_set_S128.csv}{https://github.com/SaikatBasu90/i-QER/dataset}

\subsubsection{Machine Learning Model} 

Accurate prediction of error in a quantum circuit is a difficult task due to the dynamic nature of quantum circuits and their interactions with the environment. As the feature set of the constructed training data-set and the input training data is small in size, we have applied different supervised learning methodologies such as linear regression, lasso regression, random forest, support vector regression to predict the effects of noise in a quantum circuit for a particular NISQ device, instead of deep neural networks. 

The experimental results appearing in Section \ref{section:results} below established that the support vector regression with Radial Basis Kernel is the one which captures the effects of noise on quantum hardware better than the rest of the models. Therefore, we focus on support vector regression for further computation.

The best values of penalty parameter $C$ and $\gamma$  are not known a priori, when we need to fit a given problem to a SVM model with RBF kernel. There are no generalized rules to determine these parameters. So, we have used a two stage grid-search method \cite{article} on $C$ and $\gamma$ with cross-validation on the data-set. A fine search is performed after a coarse search to find an optimal value.

\subsection{Error Influenced Binary Quantum Circuit Fragmentation}
    
Once the error in a quantum circuit is predicted to be higher than a threshold, it is fragmented into more than one partition or sub-circuit. It helps in reducing the errors in each of the sub-circuits. A quantum circuit can be represented by a tensor network $(G,A)$ consisting of a graph $G=(V,E)$ having $n$ edges and a collection of tensors $A=\{A(v):v \in V\}$ as described in Section \ref{section fragment}. A cut partitions the vertices of a graph into two disjoint subsets. Any cut determines a set of edges that have one endpoint in each subset of the partition. The number of such edges is called cut-size. A graph has $O(n!)$ combinatorial search space for all possible combination of cuts. Let us consider $K$ to be the maximum allowable cut size, and we plan to cut the circuit into two sub-circuits. Then the possible cut finding reduces to $O{n \choose  K}$. While performing our experiments we assume, $K\leq2$. Hence, the cut finding has been reduced to $O(n^{2})$.

Our objective is to reduce the error while minimizing the number of fragments. We want to cut the circuit into two circuit partitions with low correlation or entanglement between them. Cutting the circuit into two sub-circuits, makes it a special case where $K$, the number of qubits communicating between the sub-circuits becomes the upper bound of entanglement or correlation between them. In order to ensure low entanglement, a user can choose to impose a constraint on the value of $K$. Choosing a lower value of $K$, can significantly reduce the classical reconstruction cost, whereas on the other hand it can reduce the scalability of the method when it comes to dense circuits. We cut the quantum circuit if and only if there is a cut, which satisfies the constraint on cut-size. In order to reduce the error in each partition, our approach towards fragmentation is to cut the circuit in such a way that each of the partitions share approximately equal errors. Therefore, we have applied our \textit{Error Influenced Binary Cut Selector algorithm} based on binary search on the predicted error of the quantum circuit, to select a cut, so that probable error rate of each sub-circuit can reach below the threshold with fewer recursive partitioning.

\begin{algorithm}
\SetAlgoLined  
\KwInput{ $c$, number of possible cuts; all the cuts; partitions}
\KwOutput{Index of selected cut}

\tcp{It takes all the cuts with cut-size $\leq K$, where $K$ is the user-specified maximum allowable cut size}

\tcp{Predict errors of both the partitions for each cut}
\For{i=1 to c}{
  
  \tcp{Predict Errors in each of the partitions using ML Based Prediction system}
  $E_{P_{1}}=$ machine learning based prediction ($Partition_{i1}$ )\;
  $E_{P_{2}}=$ machine learning based prediction ($Partition_{i2}$ )\;
\tcp{Calculate absolute difference of error between the partitions for each cut}  
    $Distance_{i}$= $|E_{P_{1}}$-$E_{P_{2}}|$\;
}
\tcp{Find the minimum $Distance_{i}$, for $1\le i \le c$  }
\emph{Index} = index of cut at min($Distance_{i}$)\;

\KwRet{Index}

 \caption{Error influenced binary cut selector}
 \label{EIBCS}
\end{algorithm}

Algorithm \ref{EIBCS} considers each cut along with their corresponding partitions and predicts their corresponding errors. Let us assume a cut with index $i$, having two partitions $Partition_{i1}$ and $Partition_{i2}$. Algorithm 1 invokes the machine learning based prediction system to predict the errors in each of the two partitions. Further, it estimates how close these two partitions are with respect to the errors, i.e., $Distance_{i}$ is the absolute difference in error for the partitions $i1$ and $i2$. The algorithm returns the index of the cut whose $Distance$ value is minimum to have smaller height of recursive fragmentation.    

\section{Experimental Results and Analysis}

\subsection{Experimental Setup}
\label{section:results}
In this section, we describe the experimental data at each step. While the tool can be applied on any quantum hardware, we have carried out our experiments on the 15-qubit IBM Melbourne device. First, we need to identify the number of shots for which the entire experiment will be performed to ensure a consistent statistical error. Theoretically, it should be infinite to achieve a minimum statistical error. In reality, we have performed a search on the shots to identify the best performing shot with respect to the statistical errors in a quantum circuit. We have applied polynomial linear regression on the data collected from the benchmark circuits, when executed on 15-qubit IBM Melbourne device. 

\begin{figure}[!ht]
\centering
\includegraphics[scale=0.35]{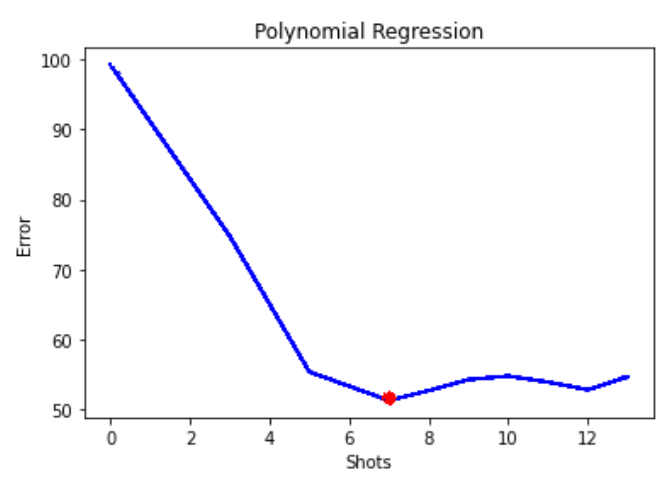}
\caption{Performance of benchmark circuits executed on 15 qubit IBM Melbourne hardware with different shots ($\log_{2}{(shots)}$) plotted in the X-axis and their corresponding mean absolute error (\%) plotted in the Y-axis }
\label{shot_vs_error}
\end{figure}

Fig. \ref{shot_vs_error} shows the error behavior of quantum circuits with respect to shots or iterations. The $x-$axis of the plot gives the index of the shot number in log-scale, i.e., the number of shots is $2^{x}$, and the $y-$axis denotes the mean of the  absolute error in the circuits run for the corresponding number of shots. We have drawn a polynomial line of best fit using linear regression to identify a better performing shot. The experimental results show that for the circuits considered, these have better performance with respect to the errors for the shot number $2^{7}=128$ marked as a red point on the plot.

\paragraph{Training data construction for the machine learning models:}

We have executed 75 different benchmark circuits from QASMBench \cite{li2020qasmbench} and Revlib \cite{4539430} on the 15-qubit IBM Melbourne device with the specified shot, to collect the training data.  QASMBench gives us a low-level, easy-to-use quantum benchmark suite, which consolidates commonly used quantum routines and algorithms from a variety of domains, and Revlib provides us with  a large database of different reversible functions.

Depending on the size of the circuit (the number of qubits), the circuits from the benchmarks can be divided into two categories. Small-scale circuits having number of qubits ranging from 2 to 5, and medium scale circuit having number of qubits ranging from 6 to 15. We have not considered large scale circuits, that is number of qubits greater than 15, as we have used 15-qubit quantum hardware for our experiments.

\begin{table}[!ht]
\scriptsize
\centering
\caption{Few important small-scale circuits from QASMBench \cite{li2020qasmbench} and Revlib \cite{4539430} used as training data}  
\begin{tabular}{|c|c|c|c|c|c|}
\hline
\textbf{Benchmark} & \textbf{Qubits} & \textbf{Gates} & \textbf{CNOT} &\textbf{Depth} & \textbf{Ref} \\ \hline

\texttt{grover}   &   3 & 22 & 2 & 14 & \cite{li2020qasmbench}  \\ \hline
\texttt{inverseqft}  & 4 & 8 & 0 & 5 & \cite{li2020qasmbench} \\ \hline
\texttt{qft}   &   4 & 12 & 0 & 8 & \cite{li2020qasmbench} \\ \hline

\texttt{Teleportation}   & 3 & 8 & 2 & 6 & \cite{li2020qasmbench} \\ \hline
\texttt{toffoli}  & 3 & 18 & 6 & 13 & \cite{li2020qasmbench} \\ \hline

\texttt{wstate} &  3 &  30 & 9 & 14 & \cite{li2020qasmbench} \\ \hline
\texttt{4mod5} &  5 &  21 & 11 & 14 & \cite{4539430} \\ \hline
\texttt{Peres} &  3 &  4 & 1 & 4 & \cite{4539430} \\ \hline
\texttt{Tof$_{dob}$} &  4 &  10 & 4 & 8 & \cite{4539430} \\ \hline

\end{tabular}
\label{tab:small}
\end{table}

\begin{table*}[!ht]
\centering\scriptsize
\caption{Few important medium-scale circuits from QASMBench \cite{li2020qasmbench} and Revlib \cite{4539430} used as training data}
\begin{tabular}{|c|c|c|c|c|c|}
\hline
\textbf{Benchmark} & \textbf{Qubits} & \textbf{Gates} & \textbf{CNOT} & \textbf{Depth} & \textbf{Ref} \\ \hline
\texttt{DNN}  & 8 & 1200 & 384 & 238 & \cite{li2020qasmbench} \\ \hline
\texttt{adder n10}  & 10 & 142 & 65 & 26 & \cite{li2020qasmbench} \\ \hline

\texttt{BVS n14}  & 14 & 41 & 13 & 16 & \cite{li2020qasmbench} \\ \hline

\texttt{multiply}  & 13 & 98 & 40 & 57 & \cite{li2020qasmbench}\\ \hline
\texttt{qaoa}  & 6 & 270 & 54 & 146 & \cite{li2020qasmbench} \\ \hline

\texttt{simons}  & 6 & 44 & 14 & 8 & \cite{li2020qasmbench} \\ \hline

\texttt{Ham7} &  7 &  62 & 31 & 37 & \cite{4539430} \\ \hline

\texttt{multipler} &  15 &  574 & 246 & 51 & \cite{4539430} \\ \hline

\end{tabular}
\label{tab:medium}
\end{table*}

In Tables \ref{tab:small} and  \ref{tab:medium}, a few important training circuits are shown. Table \ref{tab:small} contains only small scale circuits, and Table \ref{tab:medium} has medium scale circuits.

Here, we have considered the output in terms of the probability (in 0-100 scale) of the marked output states. We have compared the output probability distribution with its corresponding ideal simulation output by using the IBMQ 32-qubit QASM simulator. Then we have computed the corresponding error as the mean absolute error, $E_{mean}=\frac{1}{N}\sum_{i=1}^{N}|y_{s}-y_{q}| $ and root mean square error $E_{rmse}=\sqrt{\frac{1}{N}\sum_{i=1}^{N} (y_{s}-y_{q})^{2}}$. Here $y_{s}$ and $y_{q}$ are the probability distributions of the marked output states generated by the ideal simulator and the noisy quantum hardware respectively, and $N$ is the number of  marked output states. In the worst case, $N=2^{n}$, where $n$ is the number of qubits in the circuit. 
We have to compare two count based output probability distributions generated by the noisy quantum hardware and the ideal simulator. Another way of comparing them is to estimate the Hellinger distance \cite{Hellinger+1909+210+271}, which provides the Hellinger fidelity of the output states.

For two discrete probability distributions, ${\displaystyle P=(p_{1},\ldots ,p_{k})}$  and  ${\displaystyle Q=(q_{1},\ldots , q_{k})}$, their Hellinger distance is defined as:

\begin{equation}
    H(P,Q)=\frac{1}{\sqrt{2}}\sqrt{\sum_{i=1}^{k} (\sqrt{p_{i}}-\sqrt{q_{i}})^{2}}
\end{equation}

Hellinger fidelity is written as $(1-H^{2}(P,Q))^{2}$, which is equivalent to standard classical fidelity.
These methods to quantify error have the same worst case complexity of $O(2^{n})$, where $n$ is the number of qubits but $E_{mean}$ and $E_{rmse}$ are simpler to compute in general.

\paragraph{Performance of machine learning models:} 
Now that we have our data-set is ready, we need to train different machine based models. To perform testing on the machine learning based prediction models we have split the constructed data-set into two parts with 80\% as training data and the rest 20\% as test data. 

For Support Vector Regression, we need to choose two hyper-parameters, as mentioned in Section 3.1. For choosing $C$ and $\gamma$, we have applied the grid search \cite{article}. At first a coarse search with  $C$=1, 1000, 2000, \ldots , 50000; and  $\gamma$=1, 2, 3 is performed to identify the near to optimal values. After that, a fine search with $C$=10, 20, 30, \ldots , 1000 and $\gamma$=0.001, 0.01, 0.1, 1 is done to identify the optimum values.

The performance of the various machine learning models trained on thus constructed data-set is assessed with respect to the output generated by the models, when they are verified on test data. Root Mean Square Error (RMSE) is taken as a performance measure in this work. The coefficient of determination known as the $R^{2}$ score, on test data has also been evaluated for all the models. $R^{2}$ is defined as follows: 
\begin{equation}
R^{2}=1-\frac{SS_{R}}{SS_{T}}    
\end{equation}
 where  $SS_{R}$ is the sum of squares of the residuals or the prediction error and 
$SS_{T}$ is the total sum of squares.
Let us consider an example of the first circuit (Shor's algorithm) in Table \ref{section:results}, where the actual error of the circuit is 64.343 and our prediction system has predicted it to be 59.210. Here, $SS_{R}=(64.343-59.210)^{2} =5.133^{2}=26.347$  and $SS_{T}=64.343^{2}=4140.021$. Hence the $R^{2}=1-\frac{26.347}{4140.021}=0.993$.  
The best possible $R^{2}$ score is 1 and it may also take negative value when the model is arbitrarily worse.

\begin{figure}[!ht]
\centering
\includegraphics[scale=0.35]{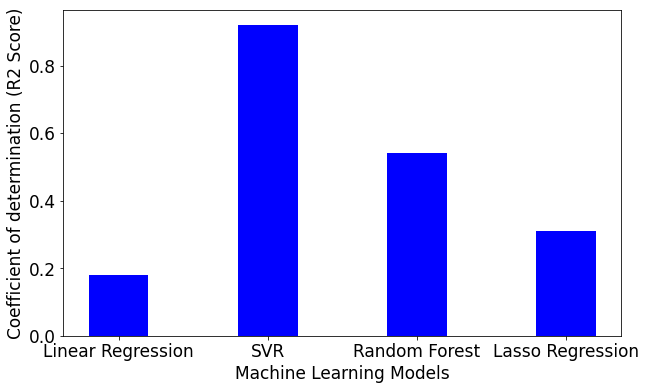}
\caption{Comparison of $R^{2}$ score for the machine learning models on our test data: Linear Regression, Lasso Regression, Random Forest and Support Vector Regression (SVR)  }
\label{r2 score}
\end{figure}

We have trained four supervised learning models: linear regression, lasso regression, support vector regression and random forest. The performance of these models on test data with respect to predicted
$R^{2}$ scores is given in Fig.\ref{r2 score} with SVR out-performing the other models.


\begin{table*}[!ht]
\footnotesize
\centering
\caption{Performance of machine learning models on test  data}
\begin{tabular}{c|c|c|c|c|c}

ML Models & RMSE & MSE & Mean Error & $R^{2}$ Score & Adjusted $R^{2}$\\
    \midrule
    Linear Regression & 31.61 & 999.51 & 26.40 & 0.18 & -0.02\\
    Lasso Regression & 29.03 & 842.93 & 24.95 & 0.31 & -0.019\\
    Random Forest & 23.67 & 560.44 & 18.19 & 0.54 & 0.32\\
    Support Vector Regression & \textbf{10.079} & \textbf{101.59} & \textbf{8.42} & \textbf{0.92} & \textbf{0.88}\\
\end{tabular}
\label{tab:performance}
\end{table*}


A detailed analysis based on the performance of the machine learning models on benchmark test circuits is given in Table \ref{tab:performance} (best values written in bold). 
We have compared machine learning models with respect to root mean squared error (RMSE), mean squared error (MSE), mean error, $R^{2}$ and adjusted $R^{2}$ score. It is observed that support vector regression (SVR) model performs better than the other commonly used supervised learning models.  The mean absolute training error for the support vector regression is 1.2305\%. The coefficient of determination ($R^{2}$) for the training data is 0.98. Hence, we have used the SVR model in our proposed prediction system $i$-QER. In order to avoid huge classical reconstruction cost, we have imposed a constraint on $K$ while performing the experiments. We have chosen $K\leq2$, for all the further experiments.

\subsection{Experimental Results}

\paragraph{Quantum circuit for validating the proposed tool $i$-QER:} We run our tool on a quantum circuit  before executing it on the real quantum hardware, say 15-qubit IBM Melbourne device. Further, we also verify the results of $i$-QER with that from the actual quantum hardware.    

\begin{figure}[!ht]
\centering
\includegraphics[scale=0.5]{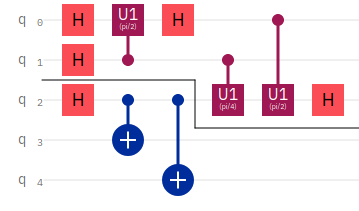}
\caption{Implementation of error influenced binary circuit fragmentation on 5 qubit Shor's algorithm }
\label{fragmentation}
\end{figure}

Fig. \ref{fragmentation} is a circuit for 5-qubit Shor's algorithm. We execute the circuit in the proposed tool. First, let us consider the threshold value for absolute error to be 50\%. Our prediction model has predicted the absolute error of the circuit to be 59.210\%. As it is higher than the predetermined threshold value, the circuit has been fragmented. We have considered the allowable cut-size, $K\leq2$. So, we have applied \textit{Error Influenced Binary Cut Selector} algorithm to cross check all possible cuts with $K\leq2$. The algorithm selected the cut with an absolute error difference of 15.94\% between the partitions, on the qubit $q2$. The detailed cut selection technique is demonstrated in Appendix \ref{appendix}. We cut the circuit as shown in Fig. \ref{fragmentation} into two circuit partitions.

We again apply the prediction model to the circuit partitions to predict their errors. As per the prediction system, sub-circuit 1 has absolute error of 10.05 and sub-circuit 2 has an absolute error of 25.99. The predicted errors are below the predetermined threshold, so these two can be directly executed on the actual hardware. While verifying, these are found to have 14.511 and 39.06 absolute error respectively. After reconstruction of the full evaluation of the circuit from the outputs of its sub-circuit partitions, we get an absolute error of 47.90. The absolute error in the circuit has been reduced by 26.129\%. The fidelity of the output state is 0.3516 when executed directly on the quantum hardware (15-qubit IBMQ Melbourne). The fidelity of the output is 0.521 when executed in our proposed technique. 

\begin{figure}[]
\centering
\includegraphics[scale=0.4]{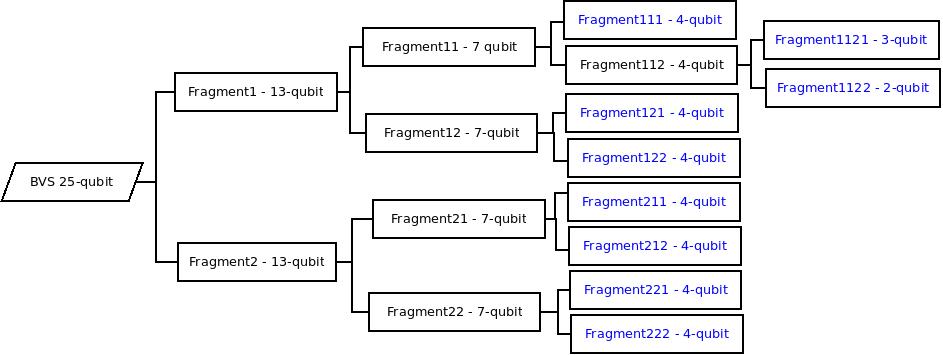}
\caption{25-qubit Bernstein-Vazirani Search (BVS) circuit fragmentation tree when executed on $i$-QER. Leaf nodes are the circuits which are fed  to the quantum hardware}
\label{BVSfragmentation}
\end{figure}

Now, let us consider a large quantum circuit of 25-qubit Bernstein-Vazirani algorithm \cite{doi:10.1137/S0097539796300921} with the same threshold. This circuit cannot be  executed directly on the  15-qubit IBMQ Melbourne device. We have considered $K\leq2$, while finding the desired cut for fragmentation.  Hence it is fragmented into two partitions, Fragment1 and Fragment2, each with 13 qubits. However, the predicted error for both the 13-qubit partitions, are found to be 100\%. Thus we need further bi-partitioning of each of the 13-qubit partitions to reduce error. Had we applied the state-of-the-art fragmentation techniques on the circuit for Bernstein-Vazirani algorithm on the 15-qubit IBMQ Melbourne device, it would not cut the circuit any further, as they can be directly executed on the quantum hardware. In our case, Fragment1 would be partitioned again into Fragment11 and Fragment12, then Fragment2  partitioned into Fragment21 and Fragment22. The predicted error for the each of the partitions  Fragment12, Fragment21 and Fragment22 is 79.115\%, and for Fragment11 is  88.72\%. So all the partitions are further partitioned as follows: Fragment11 has partitions Fragment111 and Fragment112; Fragment12 has partitions Fragment121 and Fragment122; Fragment21 has partitions Fragment211 and Fragment212; and Fragment22 has partitions Fragment221 and Fragment222. Each of these partitions has 4 qubits. The predicted error for the partition Fragment112 is 60.996\% while all the other partitions has error of 41.069\%. Thus only Fragment112 requires further split and others can be fed to the quantum hardware. Fragment112 is further fragmented into 3-qubit Fragment1121 and 2-qubit Fragment1122. The predicted error for these two are 30.85\% and 16.91\% respectively. The Fragmentation tree for the entire circuit is shown in Fig.~ \ref{BVSfragmentation}. Error after reconstruction for the entire circuit is 69.5312\%. So, we have 30.4688\% reduction in absolute error compared to earlier approaches.              

This 25-qubit Bernstein-Vazirani Search (BVS) circuit has 24 CNOT gates, which can be considered as edges of the circuit graph. The proposed cut finding algorithm takes 149 seconds to find the desired cuts, and the classical reconstruction of the sub-circuits takes 131 seconds. These proposed numerical simulations has been performed on Python 3.7, with processor Intel(R) Core(TM) i5-8250U CPU \@ 1.60GHz, RAM 8.00 GB, and 64-bit Linux Ubuntu 20.04 operating system.

\begin{table}[h!]
\footnotesize
  \centering
  \caption{Quantum circuits executed in $i$-QER and verified with corresponding actual hardware results}
  \renewcommand{\arraystretch}{1.2}
  \begin{tabular}{|p{1.5cm}|c|c|c|c|c|c|c|c|}
    \hline
    \multirow{2}{2cm}{Circuit} & \multicolumn{2}{c|}{Actual Execution} & \multicolumn{2}{c|}{Predicted Error} & Error After & Hellinger \\
    \cline{2-5}
    & Full circuit & Error in Fragments & Full circuit & Fragments &  Reconstruction & Fidelity\\
    \hline
    Shor's  &Error=64.84 & Fragment1=14.511 & 59.210 & Fragment1=10.05   & 47.90 & 0.52\\ 
     algorithm &Fidelity=0.35 &Fragment2=39.06 & & Fragment2=25.99 & & \\ \hline
    
    BVS 7qubit & Error=96.09 & Fragment1=22.656 & 86.48 & Fragment1=22.077 & 60.72 & 0.40 \\ 
    
    &Fidelity=0.03 & Fragment2=49.22 & & Fragment2=42.656 & & \\\hline
    
    $4mod5-v0_20$ &Error=82.81 & Fragment1=36.75 & 70.867 & Fragment1=32.750 & 66.20 & 0.34 \\ 
    &Fidelity=0.17 & Fragment2=46.562  & & Fragment2=41.727 & & \\\hline
    
    Toffoli  &Error=62.50 & Fragment1=29.6875 & 68.439 & Fragment1=24.366 & 39.135 & 0.61\\ 
    double &Fidelity=0.37 & Fragment2=13.4375 & & Fragment2=5.487 & & \\\hline
    
     &  & Fragment1=100  &  & Fragment1=100  &  & \\ 
     & & Fragment2=100 & & Fragment2=100 & & \\
     & & Fragment11=89.84 & & Fragment11=88.72 & & \\
     &  & Fragment12=77.34 & & Fragment12=79.11 & & \\
     & Can not & Fragment21=80.46 & & Fragment21=79.11 & & \\
     & execute & Fragment22=82.03 & & Fragment22=79.11 & & \\
     & on a & Fragment111=45.31 & & Fragment111=41.06 & & \\
    BVS 25-qubit & 15-qubit & Fragment112=63.28 & 100 & Fragment112=60.99 & 69.5312 & 0.31 \\
     & Quantum & Fragment121=38.28 & & Fragment121=41.06 & & \\
     & Hardware & Fragment122=41.40 & & Fragment122=41.06 & & \\
     & & Fragment211=35.15 & & Fragment211=41.06 & & \\
     & & Fragment212=48.43 & & Fragment212=41.06 & & \\
     & & Fragment221=46.09 & & Fragment221=41.06 & & \\
     & & Fragment222=42.18 & & Fragment222=41.06 & & \\
     & & Fragment1121=33.59 & & Fragment1121=30.85 & & \\
     & & Fragment1122=14.06 & & Fragment1122=16.91 & & \\
    \hline
    
  \end{tabular}
\label{iQer execution}  
\end{table}

We have applied our proposed methodology on the benchmark circuits and the results of these experiments are shown in Table \ref{iQer execution}. On an average, the absolute error in those circuits has been reduced by 30.169\%. The results have also shown significant increase in the fidelity of the output states when compared to that for direct hardware execution of the full circuits.

\section{Discussion and Conclusion}

We have developed a scalable classical machine learning based system \textbf{$i$-QER} to reduce errors in a quantum circuit. Supervised machine learning models are trained using the error behavior of the benchmark quantum circuits when executed in a quantum hardware, to predict the amount of errors in an unknown quantum circuit. Further, if the predicted error is higher than a threshold, the circuit is fragmented. We have proposed \textit{Error Influenced Binary Quantum Circuit Fragmentation}, a novel error-influenced fragmentation strategy, to ensure lower errors in the fragmented circuit partitions. We iterate the same procedure till the predicted error is below the user specified threshold. In order to estimate the final evaluation of the quantum circuit, we reconstruct the output probability distributions of the circuit partitions. The proposed system successfully reduced the error in quantum circuits when we conducted experiments on different quantum circuits. The tool $i$-QER, not only reduces error, but also provides a classical control over a hybrid quantum-classical computing environment. Even if the present NISQ devices scale to a larger size, this $i$-QER tool with its error-efficient classically controlled hybrid computing approach can bring more accuracy towards realistic quantum computing applications.

A few limitations of the proposed technique in its present version need to be addressed as future work. In the experimental setup, we introduced a constraint on the cut-size, i.e., $k\leq 2$, to avoid the exponential cost of classical reconstruction. This constraint  imposes a limitation to our technique when applied on a circuit with a large number of two or three qubit gates.  
This matter along with experimentation on other quantum hardware along with circuit parameters, such as topology and order of different gate application are to be explored.
The results are promising to pave the way for further research using different machine learning or neural network models on more benchmark circuits.

\begin{acks}
There is no conflict of interest. The first author gratefully acknowledges fruitful discussion regarding machine learning techniques with Mr. Dibyendu Bikash Seal, Assistant Professor, University of Calcutta, Kolkata. 
\end{acks}

\bibliographystyle{ACM-Reference-Format}
\bibliography{sample-base}


\appendix

\section{Effect of cut-selection algorithm  on error}
\label{appendix}

Let us consider the quantum circuit from Section \ref{section:results}, shown in Fig. \ref{fragmentation}. The predicted error of the circuit is 59.210 and the fidelity is 0.35157.
Here we have predetermined the threshold to be 50. Hence, partitioning of the circuit is required. To find the cut we have used our proposed Error Influenced Binary Cut Selection Algorithm. Since we have chosen the cut-size $K\leq 2$, it  considers all such cuts. It predicts error in each of the partitions for all the cuts, and identify the cut with the minimum difference in errors between the partitions. According to our algorithm, cut \{ q0, q1, q2\}; \{ q2, q3, q4\} with minimum predicted error difference of 15.94, is chosen. 

\begin{figure}[!h]
\centering
\includegraphics[scale=0.5]{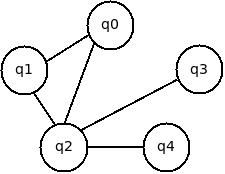}
\caption{Graph drawn for the circuit shown in Fig. \ref{fragmentation}. }
\label{graph}
\end{figure}

The circuit shown in Fig. \ref{fragmentation}, can be represented as a graph as shown in  Fig. \ref{graph}. Here, the qubits are represented as vertices. An edge between two vertices indicate a two-qubit gate operations between these two qubits. All the possible cuts along with their corresponding predicted value of error, are shown in Table \ref{cutting}. We have executed all the partitions on the 15-qubit IBMQ Melbourne device and reconstructed their output. Finally, we have estimated the error after reconstruction.    

\begin{table}[!h]
\footnotesize
  \centering
  \caption{Cuts and the predicted error for the corresponding partitions}
  \renewcommand{\arraystretch}{1.2}
  \begin{tabular}{|p{3.2cm}|c|c|c|c|c|c|c|c|}
    \hline
    \multirow{2}{2cm}{Cuts} & \multicolumn{2}{c|}{Predicted error} & \multicolumn{2}{c|}{Actual error} & Error  & Error after\\
    \cline{2-5}
    & Partition 1 & Partition 2 & Partition 1 & Partition 2 &  difference & reconstruction\\
    \hline
    \{q0\}; \{q1, q2, q3, q4\} & 2.54 & 46.112 & 5.468 & 46.875  & 41.407 & 52.95\\ \hline
    \{q1\}; \{q0, q2, q3, q4\} & 7.51 & 43.89  & 11.71 & 48.43  & 22.38 & 54.46 \\ \hline
    \textbf{\{q0, q1, q2\}; \{ q2, q3, q4\}} & 10.05 & 25.99 & 14.511 & 39.06 & \textbf{15.94} & \textbf{47.90}\\ \hline
    \{q0, q1, q2, q3\}; \{q4\} & 2.444 & 46.858 & 3.125 & 50.78 & 44.414 & 52.31\\ \hline
    \{q0, q1, q2, q4\}; \{q3\} & 2.444 & 46.58 & 4.68 & 51.56 & 44.136 & 53.82\\ \hline
    
  \end{tabular}
\label{cutting}  
\end{table}

The cut specified in bold is the one chosen by our \textit{Error Influenced Binary Cut selector} algorithm. According to our experimental results, the cut chosen by our cut selection algorithm gives us the lowest error after reconstruction.

\end{document}